\documentclass[preprint]{aastex}
\usepackage{natbib}
\usepackage{bm}
\begin{document}

\title{Fast Estimator of Primordial Non-Gaussianity from Temperature and 
Polarization Anisotropies in the Cosmic Microwave Background}

\author{
Amit P. S. Yadav\altaffilmark{1,2},
Eiichiro Komatsu\altaffilmark{3},
Benjamin D. Wandelt\altaffilmark{2,4,5}
}
\altaffiltext{1}{email: ayadav@uiuc.edu}
\altaffiltext{2}
{Department of Astronomy, University of Illinois at 
Urbana-Champaign, 1002 W.~Green Street, Urbana, IL 61801}
\altaffiltext{3}{
Department of Astronomy, University of Texas at Austin, 2511 Speedway, RLM 15.306, TX 78712}

\altaffiltext{4}{
Department of Physics,  University of Illinois at Urbana-Champaign, 1110 W.~Green Street, Urbana, IL 61801}

\altaffiltext{5}{Center of Advanced  Studies,University of Illinois at
Urbana-Champaign, 912, W.~Illinois Street, Urbana, IL 61801}

\begin{abstract}
Measurements of primordial non-Gaussianity ($f_{NL}$) open a new
 window onto the physics of inflation.
We  describe a  fast cubic (bispectrum) estimator of $f_{NL}$,
using a combined analysis of temperature  and polarization observations.
The speed of our
estimator allows us to use a sufficient number of Monte Carlo simulations
 to characterize its statistical properties in the presence of real
 world issues such as
instrumental effects, partial sky coverage, and foreground
contamination. We find that our estimator is optimal, where
optimality is defined by  saturation of the Cramer Rao bound, if noise
 is homogeneous.
Our  estimator is also computationally efficient, scaling as  $O(N^{3/2})$
compared to the $O(N^{5/2})$  scaling of the
brute force  bispectrum calculation for sky maps with $N$ pixels.
For Planck this translates into a  speed-up by factors of
millions, reducing the required computing time from thousands of years to
just hours and thus making $f_{NL}$ estimation feasible for future
surveys. Our estimator in its current form is optimal if noise is
 homogeneous. In future work our fast polarized bispectrum estimator should be
extended to deal with inhomogeneous noise in an analogous way to how the
existing fast temperature estimator was generalized. 
\end{abstract} 

\section{Introduction} 
\label{introduction}
In the  last few decades  the advances in the  observational cosmology
have led the field to its ``golden age.'' Cosmologists are beginning to
nail down  the basic cosmological parameters, and  have started asking
questions  about the  nature of  the initial  conditions  provided by
inflation~\citep{Guth81, Sato81, Linde81,Albrecht_Steinhardt82},  which apart
from solving the flatness and  horizon problem, also gives a mechanism
for  producing the  seed  perturbations for  structure formation~\citep{Guth_Pi82,Starobinsky82,
Hawking82,Bardeen_et83,Mukhanov_et92},  and
other  testable  predictions.  

The  main predictions  of  a  canonical
inflation model are: (i) spatial flatness of the observable
universe,  (ii)  homogeneity and  isotropy on  large
scales  of  the   observable  universe, (iii)  nearly  scale
invariant  and  adiabatic   primordial  density  perturbations,  and
(iv) primordial perturbations to  be very close to Gaussian. 
The cosmic microwave background (CMB) data from the Wilkinson Anisotropy
Probe (WMAP) \citep{wmap_1st_bennett}, both temperature
\citep{wmap_1st_hinshaw,wmap_2nd_hinshaw} and polarization
\citep{wmap_1st_pol,wmap_2nd_pol} anisotropies, have provided hitherto the strongest
constraints on these
predictions~\citep{nong_wmap,Peiris03,wmap_1st_spergel,wmap_2nd_spergel}. There
is no observational evidence against simple inflation models.


Non-Gaussianity from the simplest inflation models that are based on a
slowly rolling scalar field   is    very
small~\citep{Salopek_Bond90,Salopek_Bond91,Falk_et93,Gangui_et94,Acquaviva02,Maldacena03};
however, a very large class of more general models with, e.g., multiple
scalar fields, features in inflaton potential, non-adiabatic fluctuations, non-canonical kinetic terms, deviations from Bunch-Davies
vacuum, among others \cite[for a review and references therein]{BKMR_04} 
generates substantially higher
amounts of non-Gaussianity.

The amplitude of non-Gaussianity constrained from the data 
is often quoted  in terms  of a non-linearity parameter
$f_{NL}$  (defined   in  section~\ref{model}). Many efficent methods for evealuating bispectum of CMB temperature anisotropies 
exist ~\citep{KSW05,IntegratedBispectrum,SmithZaldarriaga}. 
So far, the bispectrum tests of non-Gaussianity have
not detected any significant $f_{NL}$ in temperature fluctuations mapped by
COBE~\citep{nong_bdw} and
WMAP~\citep{nong_wmap,wmap_2nd_spergel,creminelli_wmap1,creminelli_wmap2,cabella06,szapudi06}. 
 Different   models  of
inflation predict different amounts  of $f_{NL}$, starting from $O(1)$
to $f_{NL}\sim 100$, above which values have been excluded by the WMAP
data already. On the other hand,
some authors have claimed non-Gaussian signatures in the WMAP
temperature
data~\citep{hotcold,larsonwandelt,nong_cmb1,nong_cmb2,nong_cmb3}. These
signatures cannot be characterized by $f_{NL}$ and are consistent with
non-detection of $f_{NL}$. 

Currently  the constraints on  the $f_{NL}$  come from  
temperature anisotropy data alone. By  also having  the polarization
information in the  cosmic microwave background, one can improve
sensivity to  primordial fluctuations \citep{BZ04,YW05}. Although
the  experiments have alrady  started characterizing  polarization
anisotropies \citep{dasi_pol_02, wmap_1st_pol, wmap_2nd_pol,boom_ee}, the errors  are large
in comparison to temperature anisotropy. The upcoming experiments such
as Planck will characterize polarization anisotropy to high
accuracy. Are we ready to use future polarization data for  testing Gaussianity of primordial
fluctuations? Do we have a fast estimator which allows us to measure $f_{NL}$
from the combined analysis of temperature and polarization data? 

In  this paper we extend the fast cubic estimator of $f_{NL}$ from the
temperature data \citep{KSW05} and
derive a  fast way for measuring primordial non-Gaussianity using
the cosmic microwave background  temperature and polarization maps. We
construct a  cubic statistics, a cubic combination of (appropriately
filtered) temperature and polarization maps, which is specifically  sensitive to the
primordial  perturbations. This  is done  by reconstructing  a  map of
primordial perturbations,  and using that to define  our estimator. We
also show  that the inverse of  the covariance matrix  for the optimal
estimator~\citep{BZ04} is the  same as the product of  inverses we get
in the  fast estimator. Our estimator takes  only $N^{3/2}$ operations
in comparison to the full bispectrum calculation which takes $N^{5/2}$
operations. Here $N$ refers to  the total number of pixels. For Planck
$N \sim  5 \times 10^7$,  and so the  full bispectrum analysis  is not
feasible while ours is.
\section{Primordial Non-Gaussianity}
\subsection{A Model}
\label{model}
The harmonic coefficients of the CMB anisotropy $a_{lm}=T^{-1}\int d^2
{\mathbf {\hat n}}\Delta T({\mathbf {\hat n}}) Y^{\star}_{\ell m}$ can
be related to the primordial fluctuation as:
\begin{eqnarray} 
\label{phi_alm}
a_{\ell    m}^X=\frac{2b_{\ell}}{\pi}\int   k^2dk\,r^2dr\,\lbrack   \,
 \Phi_{\ell    m}(r)   \,    g^{adi}_{X\ell}(k)   +    S_{\ell   m}(r)
 \,g^{iso}_{X\ell}(k)    \,    \rbrack    \,    j_\ell(kr)+    n_{\ell
 m}\end{eqnarray} where  $\Phi_{\ell m}(r)$  and $S_{\ell m}(r)  $ are
 the harmonic  coefficients of the  primordial curvature perturbations
 and the primordial isocurvature perturbations respectively at a given
 comoving distance $r=\vert \mathbf  {r} \vert$; $g_{X\ell}(r)$ is the
 radiation  transfer  function  of  either adiabatic  or  isocurvature
 perturbations; X refers  to either T or E;  and $j_{\ell}(kr)$ is the
 Bessel function of  order $\ell$. A beam function  $b_{\ell}$ and the
 harmonic  coefficient  of noise  $n_{\ell  m}$  are instrumental  and
 observational  effects.  Eq.  (\ref{phi_alm})  is  written  for  flat
 background, but can easily be generalized.

Any   non-Gaussianity   present   in  the   primordial   perturbations
$\Phi_{\ell  m}(r)$ or  $S_{\ell  m}(r)$, can  get  transfered to  the
observed  CMB  i.e. $a_{lm}$,  via  Eq.  (\ref{phi_alm}).  Due to  the
smallness of isocurvature contribution over the curvature perturbation
\citep{Peiris03,Bean_etal_06,Trotta06}  we   will   drop  the
isocurvature  contribution from  Eq.  (\ref{phi_alm})  and  further we
will  use a  popular and  simple non-Gaussianity  model \citep{Gangui_et94,verde00,KS2001} given by
\begin{eqnarray}
\Phi(r)=\Phi_L(r)+f_{NL}\lbrack     \Phi^2_L(r)-\langle    \Phi^2_L(r)
\rangle \rbrack
\label{phing}
\end{eqnarray}
where $\Phi_L(r)$  is the linear  Gaussian part of  the perturbations, and
$f_{NL}$  is a non-linear  coupling constatnt  characterizing  the
amplitude  of the  non-Gausianity. The  bispectrum in  this model  can be
written as:
\begin{eqnarray}
\label{phi_bispec}
\langle
\Phi(k_1)\Phi(k_2)\Phi(k_3)=2f_{NL}(2\pi)^3\delta^3(k_1+k_2+k_3)
\lbrack P(k_1)P(k_2)+cycl.\rbrack.
\end{eqnarray}  
The above form  of the bispectrum is specific to  the model chosen and
so in  general the constrains on  $f_{NL}$ do not  necessarily tell us
about non-Gaussianity of other models \citep{babich04,BKMR_04}.
This model, for instance, fails completely when non-Gaussianity is
localized in a specific range in $k$ space, the case that is predicted
from inflation models with features in inflaton potential
\citep{wang00,nong_wmap,chen06}.
Even for the simplest inflation models based on a slowly rolling scalar
field, the bispectrum of {\it inflaton perturbations} yields a non-trivial
scale dependence of $f_{NL}$ \citep{Falk_et93,Maldacena03},
although the amplitude is too small to detect. On the other hand the
bispectrum of {\it curvature perturbations} contains the contribution
after the horizon exit, which is non-zero even when inflaton
perturbations are exactly Gaussian. This contribution actually follows
Eq.~(\ref{phi_bispec}) \citep{lyth05}. Curvaton models also yield the bispectrum
in the form given by Eq.~(\ref{phi_bispec}) \citep{lyth03}.

\subsection{Reconstructing Primordial Perturbations using Temperature
  and Polarization Anisotropy}

Reconstruction of primordial perturbations from the cosmological data
allows  us to  be  more  sensitive  to primordial non-Gaussianity, which
is important because non-Gaussianity in the
cosmic microwave  background  data does not necessarily imply  the presence of
primordial non-Gaussianity. The  method of reconstruction from 
temperature data is described in \citep{KSW05} and that from the
combined analysis of temperature and polarization data is described in \citep{YW05},  where we  reconstruct  the perturbations  $\Phi(r)$
using an operator $O_{\ell}(r)$. These operators are given by:

\begin{eqnarray} \left( \begin{array}{c} O_{\ell}^T(r)\\O_{\ell}^E(r)\\ \end{array} \right) = \left( \begin{array}{ccc} 
C_{\ell}^{TT}&C_{\ell}^{TE}\\C_{l}^{TE}&C_{l}^{EE}\\\end{array}
\right)^{-1}\,     \left(     \begin{array}{c}     \beta_{\ell}^T(r)\\
\beta_{\ell}^E(r)  \\   \end{array}  \right),  \label{3}\end{eqnarray}
where

\begin{eqnarray} C_{\ell}^{XY}&\equiv&\langle  a_{\ell m}^X a_{\ell m}^{*Y}\rangle
 =\frac  {2b^X_{\ell}b^{Y}_{\ell}}{\pi}   \int  k^2  dk  \,P_{\Phi}(k)
\,g_{X\ell}\,g_{Y\ell}(k) + N^{XY}_{\ell}
\end{eqnarray}

\begin{eqnarray}
\label{beta}
 \beta_{\ell}^X(r)&   \equiv   &\langle   \Phi_{\ell   m}(r)   a_{\ell
m}^{*X}\rangle  =\frac {2b^X_{\ell}}{\pi}  \int  k^2 dk  \,P_{\Phi}(k)
\,g_{X\ell}(k)\, j_{\ell}(kr),
\end{eqnarray}
$P_{\Phi}(k)$  is  the  power  spectrum of  the  primordial  curvature
perturbations,   and  $N^{XY}_{\ell}=\langle  a^X_{\ell   m}  a^{\star
Y}_{\ell m}\rangle$  is the noise covariance  matrix. The primordial perturbation
then can be estimated as:
\begin{eqnarray} \hat{\Phi}_{\ell m}(r) = \sum_{X=T,E}{O_{\ell}^X(r)}\,a_{\ell m}^{X}\label{6} \end{eqnarray}
Figure 1 shows the improvement in reconstruction due to additional information from the CMB polarization.

\begin{figure*}[h]
\includegraphics[height=8cm,                  angle=90]{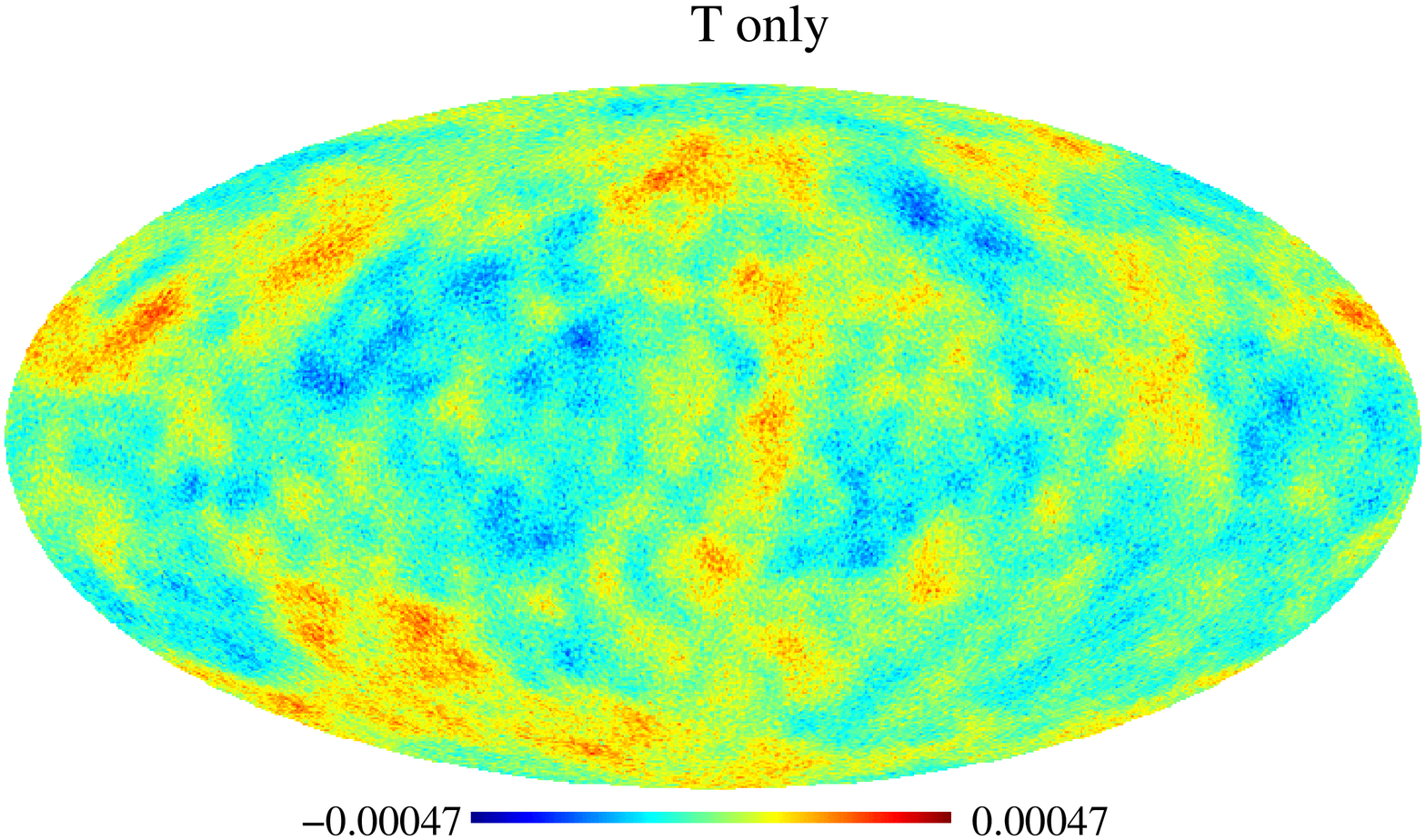}
\includegraphics[height=8cm, angle=90]{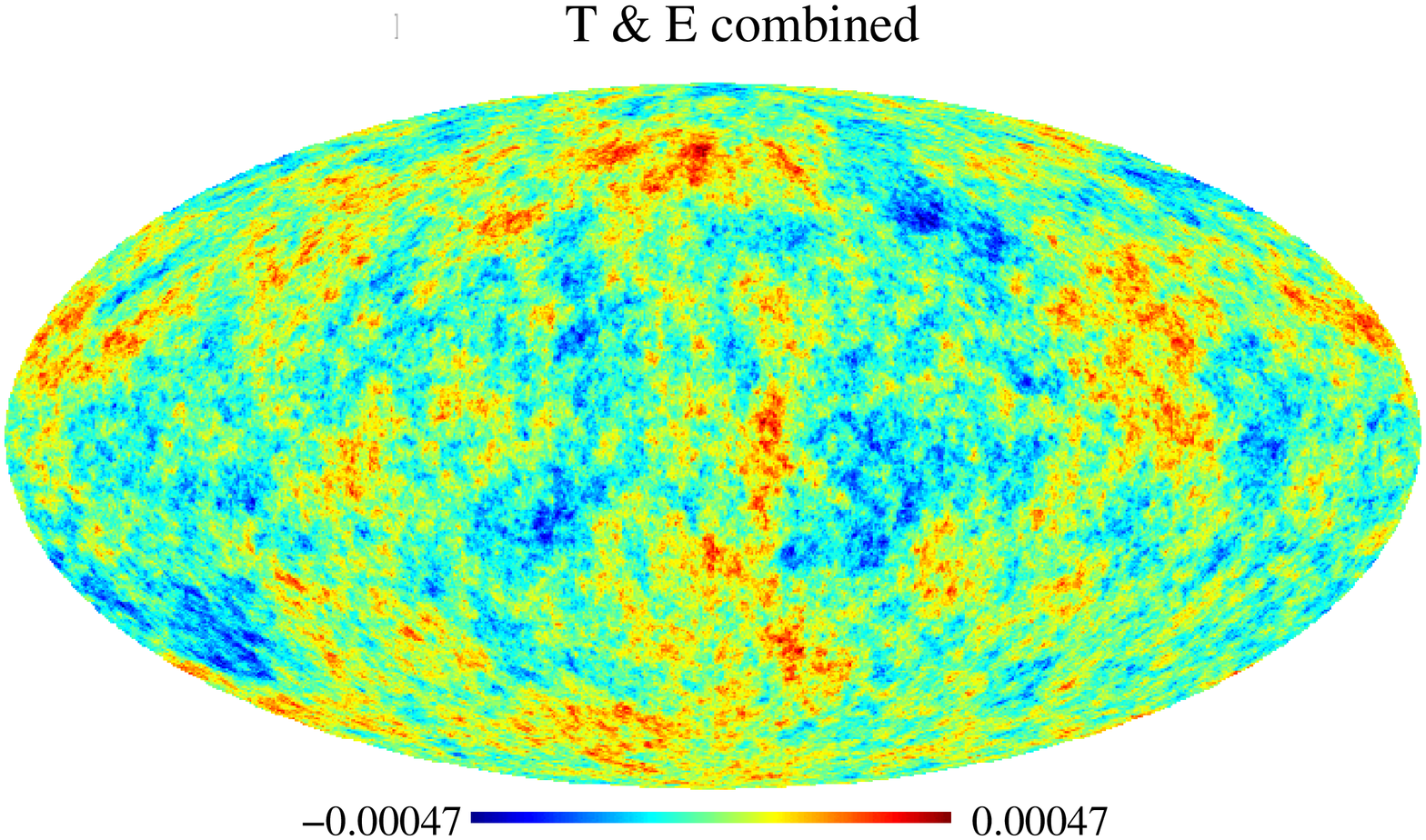}
\caption{Reconstructed   primordial   perturbation   map using   only
temperature  information  (left),  and,  using  both  temperature  and
polarization information (right). Perturbations are reconstructed at the surface of last scattering.\label{fig1}}
\end{figure*}
\subsection{Fast Cubic Statistics}
We can construct a quantity  $\hat{S}_{prim}$ that is analogous to the
KSW    fast   estimator of $f_{NL}$ from temperature data~\citep{KSW05}
but    generalize it    to include 
polarization data as well. 
This  quantity has a  simple interpretation in  terms of
the tomographic  reconstruction of the  primoridal potential described
in~\citep{KSW05,YW05}. It is  the radial integral of a  cubic combination of
the scalar potential reconstructions, the B terms with the analogous A
term.

As in \citep{KSW05}, we form a cubic statistics given by
\begin{eqnarray}
\label{s_prim}
\hat{S}_{prim}=\frac{1}{f_{sky}}\int     r^2dr     \int     d^2\hat{n}
B(\hat{n},r) B(\hat{n},r)A (\hat{n},r)
\end{eqnarray} 
where
\begin{eqnarray}
\label{B}
B(\hat{n},r)\equiv           \sum_{ip}\sum_{lm}(C^{-1})^{ip}a^{i}_{\ell
m}\beta^p_{\ell}(r)Y_{\ell m}(\hat{n}),
\end{eqnarray}
\begin{eqnarray}
\label{A}
A(\hat{n},r)\equiv           \sum_{ip}\sum_{lm}(C^{-1})^{ip}a^{i}_{\ell
m}\alpha^p_{\ell}(r)Y_{\ell m}(\hat{n}),
\end{eqnarray}
$\alpha^p_{\ell}=\frac  {2b^p_{\ell}}{\pi} \int k^2  dk g_{X\ell}(k)\,
j_{\ell}(kr)$, and $f_{sky}$ is a  fraction of sky. Index $i$ and $p$
can  either be  $T$ or  $E$. We  find (see  Appendix A)  that $\langle
\hat{S}_{prim} \rangle$ reduces to
\begin{eqnarray}
\langle  \hat{S}_{prim}  \rangle  =\frac{1}{f_{sky}}\sum_{i  j k  p  q
 r}\sum_{2 \le \ell_1\le  \ell_2\le \ell_3}  \frac{1}{\Delta_{\ell_1 \ell_2 \ell_3}}f_{NL}  B^{p q  r,prim}_{\ell_1
 \ell_2      \ell_3}(C^{-1})^{ip}_{\ell_1}(C^{-1})^{j      q}_{\ell_2}
 (C^{-1})^{k  r}_{\ell_3}B^{i  j   k,prim}_{\ell_1  \ell_2  \ell_3  },
 \end{eqnarray} where $\Delta_{\ell_1  \ell_2 \ell_3}$ is: 1 when  $\ell_1 \neq \ell_2
\neq \ell_3$, 6 when $\ell_1 =  \ell_2 = \ell_3$, and 2 otherwise; $B^{p q r,prim}_{\ell_1 \ell_2  \ell_3}$ is the
 theoretical bispectrum for $f_{NL}=1$, and is given by:

\begin{eqnarray}
B^{p  q r,prim}_{\ell_1 \ell_2  \ell_3}= I_{\ell_1  \ell_2 \ell_3}\int
r^2dr      \lbrack     \beta^p_{\ell_1}(r)\beta^{q}_{\ell_2}(r)\alpha^
{r}_{\ell_3}(r)                                                       +
\beta^{r}_{\ell_3}(r)\beta^{p}_{\ell_1}(r)\alpha^{q}_{\ell_2}(r)      +
\beta^{q}_{\ell_2}(r)\beta^{r}_{\ell_3}(r)\alpha^{p}_{\ell_1}(r)
\rbrack \end{eqnarray} where
\begin{eqnarray} 
I_{\ell_1    \ell_2    \ell_3}=2   \sqrt{\frac{(2\ell_1    +1)(2\ell_2
+1)(2\ell_3  +1)}{4\pi}}\left(\begin{array}{ccc}  \ell_1  &  \ell_2  &
\ell_3\\ 0 & 0 & 0\end{array}\right)
\end{eqnarray}

Since $\langle  \hat{S}_{prim} \rangle$ is proportional  to $f_{NL}$,
an unbiased estimator of $f_{NL}$ can be written as:
\begin{eqnarray}
\label{fnl_estimate}
\hat{f}_{NL} = \frac{\hat{S}_{prim}}{\left\{  \sum_{i j  k  p q
r}\sum_{2 \le \ell_1\le  \ell_2\le  \ell_3} \frac{1}{\Delta_{\ell_1 \ell_2 \ell_3}} B^{p q  r,prim}_{\ell_1  \ell_2
\ell_3}(C^{-1})^{ip}_{\ell_1}(C^{-1})^{j         q}_{\ell_2}(C^{-1})^{k
r}_{\ell_3} B^{i j k,prim}_{\ell_1 \ell_2 \ell_3 }\right\}}
\label{fastestim}
\end{eqnarray}
The  most time  consuming  part  of the  calculation  is the  harmonic
transformation  necessary for Eqs.  (\ref{B}) and(\ref{A}).   The fast
estimator  defined above  takes  only $N^{3/2}$  operations times  the
number of sampling points for $r$, which  is of the order 100. Hence this
is  much faster  than  the  full bispectrum  analysis,  which scales  as
$N^{5/2}$. $N$ here is the number of pixels. In the next section we
will show that this fast unbiased estimator is also optimal, by
proving equivalence with a known optimal but slow estimator.

\begin{figure*}[h]
\includegraphics[height=6cm, angle=0]{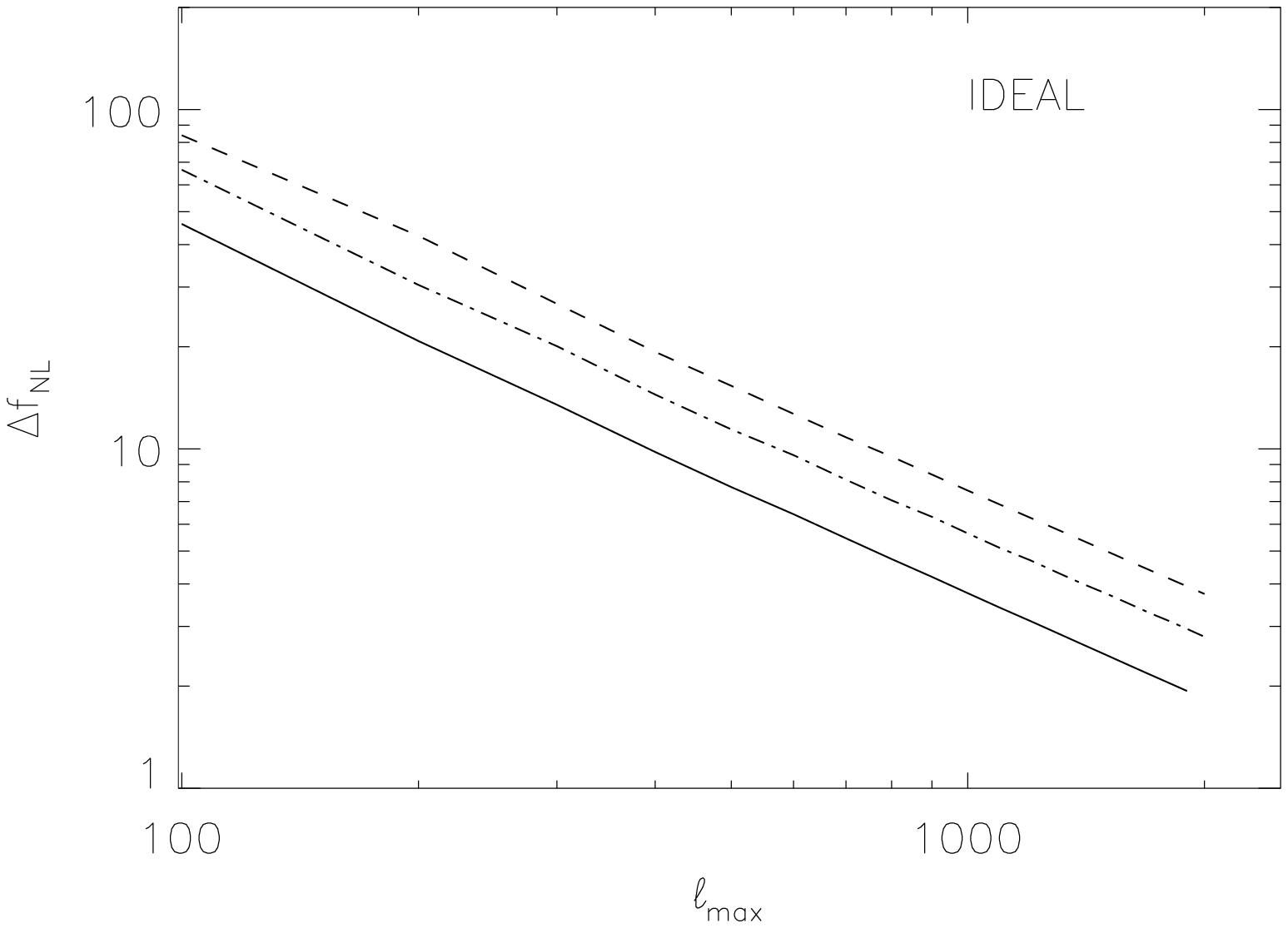}
\includegraphics[height=6cm, angle=0]{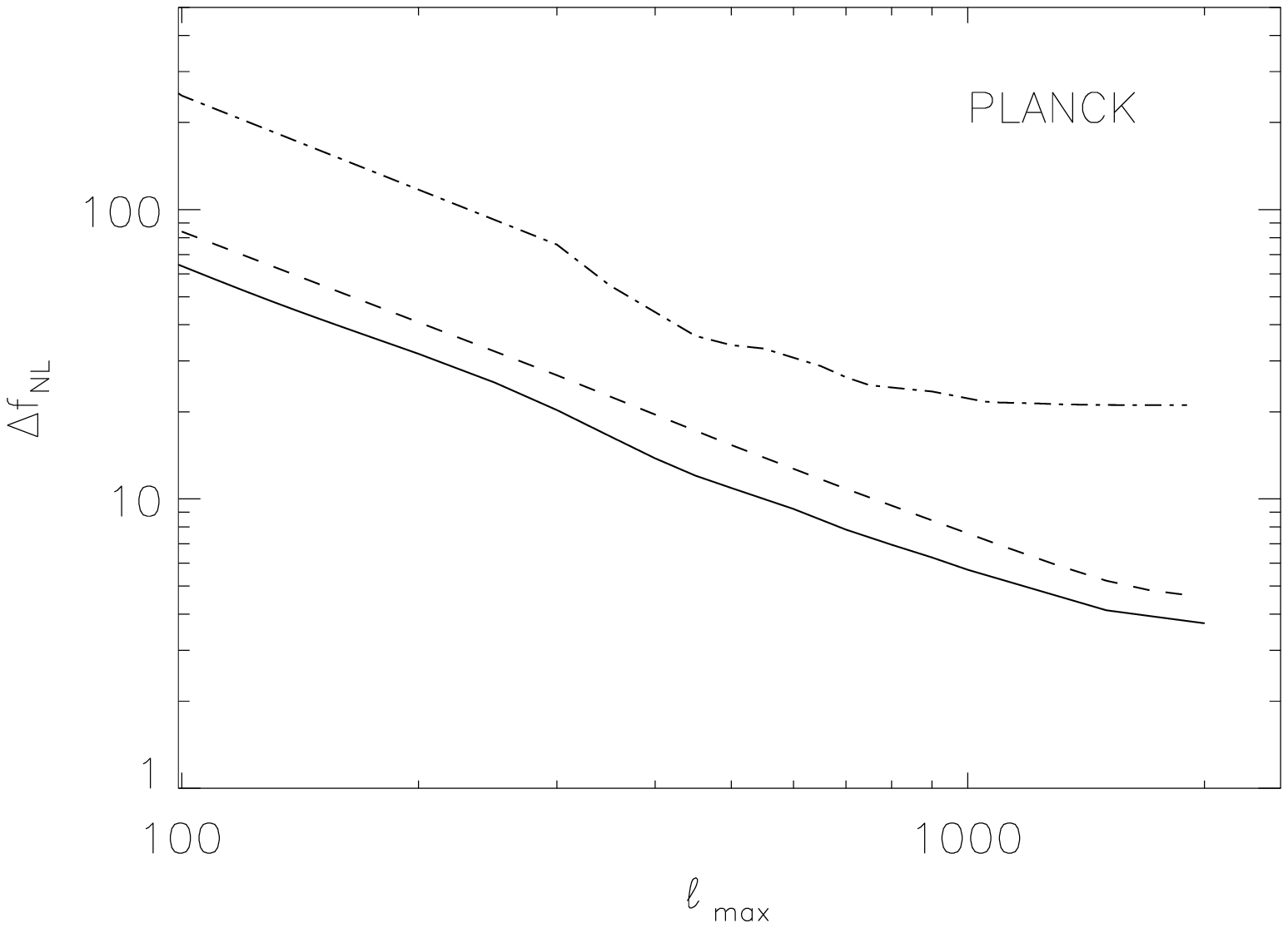}
\caption{Fisher  predictions  for  minimum detectable  $f_{NL}$ at the 1-$
\sigma$ level.  Left
panel: ideal experiment. Right panel: Planck satellite.
Solid  lines:    temperature   and  polarization
information combined.  Dashed  lines:  temperature
information only.  Dot-dashed line: polarization
information only.
  \label{fig2}}
\end{figure*}

\subsection{Optimality of the fast estimator} 
In their recent paper  Babich and Zaldarriaga \citep{BZ04} have found
an optimal  estimator for $f_{NL}$ which minimizes the expected
$\chi^2$ given by
\begin{eqnarray} 
\chi^2=\sum_{ijkpqr}\sum_{\ell_1\ell_2\ell_3}\left(
B^{ijk,obs}_{\ell_1\ell_2\ell_3}-f_{NL}B^{ijk,prim}_{\ell_1\ell_2\ell_3}\right)({\bf
Cov}^{-1})^{ijk}_{pqr}                                           \left(
B^{pqr,obs}_{\ell_1\ell_2\ell_3}-f_{NL}B^{pqr,prim}_{\ell_1\ell_2\ell_3}\right).
\end{eqnarray}
This optimal estimator is
\begin{eqnarray}
\hat   f_{NL}=\frac{\sum_{i  j   k  p   q  r}\sum_{\ell_1\ell_2\ell_3}
 B^{ijk,obs}_{\ell_1\ell_2\ell_3}({\bf              Cov}{-1})_{ijk,pqr}
 B^{pqr,prim}_{\ell_1\ell_2\ell_3}}{\sum_{ijkpqr}\sum_{\ell_1\ell_2\ell_3}
 B^{ijk,prim}_{\ell_1\ell_2\ell_3}({\bf            Cov}^{-1})_{ijk,pqr}
 B^{pqr,prim}_{\ell_1\ell_2\ell_3}},
\end{eqnarray}
where  the  index $ijk$  and  $pqr$  runs over  $\{TTT,TTE,TEE,EEE\}$,
unlike in the  fast estimator case where $ijk$ and  $pqr$ run over all
the 8 combinations $\{TTT,TTE,TET,ETT,TEE,ETE,EET,EEE\}$.  In appendix
B  we show  that  the inverse  of  the covariance  matrix  for the  BZ
estimator is  the same as the product  of inverses we get in the fast
estimator, Eq.~\ref{fastestim}; hence our estimator is
optimal\footnote{By an analogous argument one can show that the
  temperature-only estimator in Komatsu et al. (2005) is also optimal,
  not slightly suboptimal as stated.}

\begin{figure*}[!ht]
\includegraphics[height=12cm,angle=0]{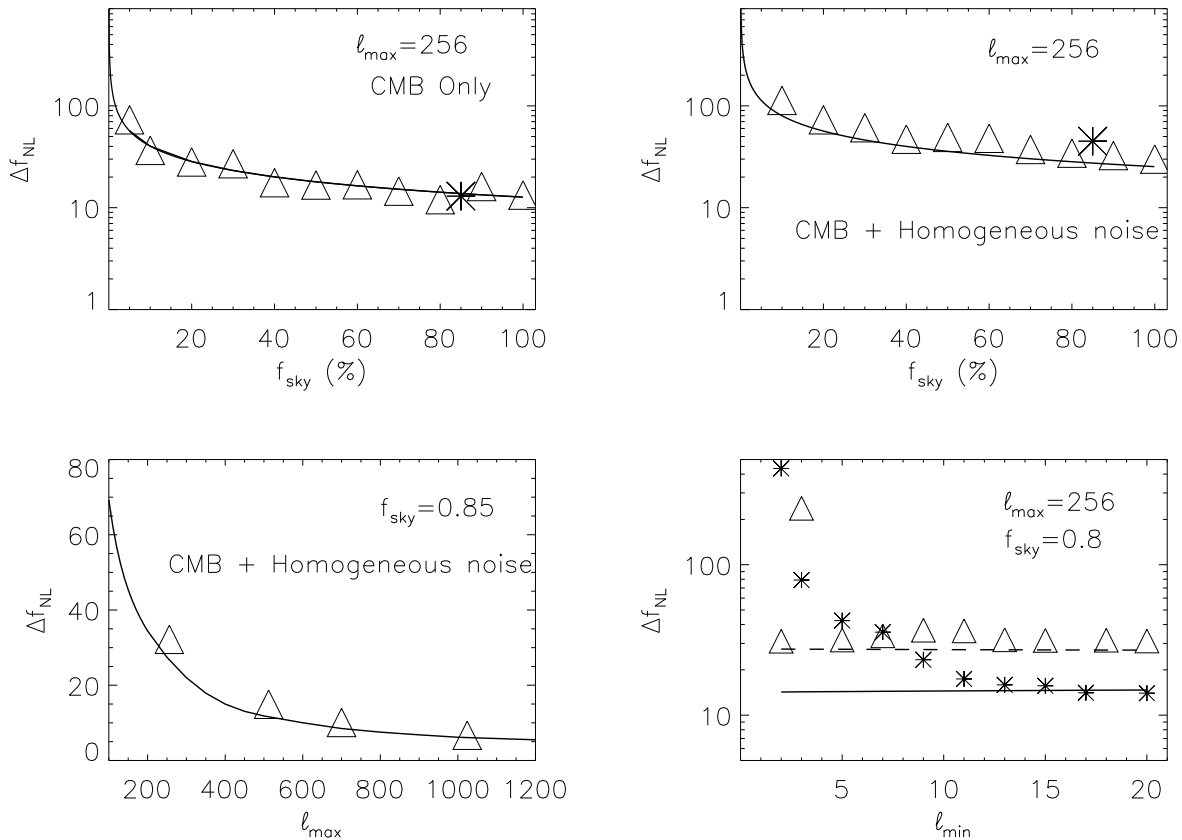}

\caption{Testing optimality of our $f_{NL}$ estimator. 
In all panels lines show the optimal Cramer Rao
bounds given by $\left( F^{-1}/f_{sky} \right) ^{1/2}$. Symbols
show the 1-$\sigma$ errors on $f_{NL}$ derived from Monte Carlo simulations.
{\bf Upper panels:} Monte Carlo errors as a
function of $f_{sky}$. 
The star shows the WMAP Kp2 mask. 
Triangles:  
straight sky cuts excluding regions of low galactic latitudes. 
The left panel shows only the effect of the mask, while 
the right panel includes homogeneous white noise and beam
smoothing at the level of the Planck satellite. 
{\bf Lower left:}
Monte Carlo errors as a function of $\ell_{max}$.
{\bf Lower right:} 
Incomplete sky coverage
causes excess variancde of the low $\ell$ modes of the polarization
bispectra. We show results as a function of $\ell_{min}$, below which
multipoles have been removed from  the analysis of polarization
bispectra in two cases: 1) stars show the  simluated errors  in the 
noiseless case; 2)  the dashed line and triangles
show a case with homogeneous noise similar to the Planck satellite. 
The variance excess due to overweighting of the polarization modes is
cleary visible in the noiseless case. This
panel demonstrates that excluding the lowest $\ell$ polarization modes from
the analysis avoids this variance excess without significant
loss of information. 
\label{fig3}}

\end{figure*}

\section{Results} 
To test optimality of our fast estimator we use 
Eqs. (\ref{fnl_estimate})  and (\ref{s_prim}) to measure $f_{NL}$ from
simulated Gaussian skies. The
error  bars on $f_{NL}$ are then derived  from  Monte Carlo  simulations  of our  fast
estimator.  The  simulated errors are compared with the Cramer Rao
bound  $(F^{-1}/f_{sky})^{1/2}$, where $f_{sky}$ is a fraction
of sky, and $F$ is the Fisher matrix given by~\citep{KS2001, BZ04}:
\begin{eqnarray}
F = \sum_{i  j  k p  q  r}\sum_{\ell_1 \le  \ell_2 \le  \ell_3}
 \frac{1}{\Delta_{\ell_1 \ell_2 \ell_3}} B^{p q r,prim}_{\ell_1 \ell_2
 \ell_3}(C^{-1})^{ip}_{\ell_1}(C^{-1})^{j    q}_{\ell_2}   (C^{-1})^{k
 r}_{\ell_3} B^{i j k,prim}_{\ell_1 \ell_2 \ell_3},
\end{eqnarray}
where $\Delta_{\ell_1  \ell_2 \ell_3}$ is: 1 when  $\ell_1 \neq \ell_2
\neq \ell_3$, 6 when $\ell_1 =  \ell_2 = \ell_3$, and 2 otherwise.  We
have  used  $\Lambda$CDM  model  with  $\Omega_c=0.26,  \Omega_b=0.04,
\Omega_{\Lambda}=0.7,  h=0.7$,  and a constant scalar spectral index $n_s=1$. Since  the
contribution to the integral in  Eqs. \ref{B} and \ref{A} comes mostly
from  the decoupling epoch  (for our  simulations $r_{dec}=13.61$Gpc),
our   integration   limits    are   $r_{min}=13425$   Mpc   and
$r_{max}=13865$ Mpc,  with the sampling  $dr=15$ Mpc. Refining the
sampling of the $r$ integral by a factor of 2 does not change our results significantly. The  results are
summarized in figure~\ref{fig3}. We separately explore the effects of
excluding foreground contaminated sky regions and noise degradation
due to the instrument. For definiteness we have used the published Planck noise amplitudes
which are described in Table 2, though we ignore the effect of the
scanning strategy and noise correlations. We shall come back to this
point in \S~\ref{sec:conclusions}.


We find that the  low $\ell$
modes of the polarization bispectrum are contaminated significantly
when $f_{sky}<1$. 
This is illustrated  by the
lower right panel  of Figure~\ref{fig3}, where we sum over all
the  $\ell$ modes  for $TTT$  contribution to the bispectrum (because  the temperature
bispectrum is not contaminated by the sky cut),  while varying $\ell_{min}$
for the other terms  ($EEE,  TTE,  TEE$). Our results show that 
one may simply remove the contamination in the
polarization bispectrum due to the
sky cut by removing $\ell\lesssim 10$ (when
$f_{sky}\sim 0.8$)  without sacrificing
sensitivity to $f_{NL}$.
The contamination appears to be less when noise is added, as
the dominant constraint still comes from the temperature data
which are insensitive to the contamination due to the sky cut.

We can explain this behaviour simply in terms of the coupling between
spherical harmonic modes induced by the sky cut. The key observation
is that the low $\ell$ polarization spectrum is very small in the
theoretical model we chose, but if it 
could be observed it would add  information to the  $f_{NL}$
estimator. Therefore, for a
noiseless experiment, an optimal estimator designed for full sky work will assign a large weight to
the low $\ell$ polarization modes. Since the low $\ell$ spectrum rises
very steeply towards higher $\ell$ a sky cut creates power leakage
from higher $\ell$ to lower $\ell$. This biases the low $\ell$
spectrum significantly and this bias is amplified by the large
coefficient  the estimator assigns to these modes. In a realistic
experiment including noise, the low
$\ell$ polarization modes are difficult to measure, since noise
dominates. Accordingly, the estimator assigns small weights to the low $\ell$
polarization modes and the sky cut has a much less prominent effect.

\subsection {Computational Speed}

Our  fast  estimator  takes  only  $N^{3/2}$
operations times the  number of sampling points for the integral over 
$r$,  which is of the
order 100. Hence this is much faster than the full bispectrum analysis
as discussed in~\citep{BZ04}, which goes as $N^{5/2}$. $N$ here is the
number  of pixels. For  Planck we  expect $N\sim  5 \times  10^7$, so
performing 100 simulations using 50 CPUs takes only 10 hours using our
fast estimator, while we estimate it would take approximately $10^3$ years to do
the brute force bispectrum calculation using the same platform.

\begin{table}
\caption{Planck noise properties assumed for our analysis\label{tbl-2}}
\begin{tabular}{|c|ccccccc|}
\hline &\multicolumn{7}{c|}{Central frequency (GHZ)}\\  & 30 & 44 & 70
& 100 & 143 & 217 & 353 \\ \hline Angular Resolution [FWHM arcminutes]
&33 & 24 & 14 & 10 &  7.1 & 5.0 & 5.0 \\ $\Delta{T} /T$ intensity $^a$
[$10^{-6}  \mu$K/K]  &2.0&2.7&4.7&2.5&2.2&4.8&14.7  \\ $\Delta{T}  /T$
polarization     (Q     and     U)$^a$    [$10^{-6}     \mu$K/K]     &
2.8&3.9&6.7&4.0&4.2&9.8&29.8\\ \hline
\end{tabular}\\
\tablenotetext{a}{Average 1$\sigma  $ sensitivity per  pixel (a square
whose  side  is  the  FWHM  extent of  the  beam),  in  thermodynamic
temperature units, achievable after 2 full sky surveys (14 months).}
\end{table}

\section{Conclusions}
\label{sec:conclusions}
 Starting with the tomographic reconstruction approach \citep{KSW05,YW05} we  have found
 a fast,  feasible,  and optimal  estimator of $f_{NL}$, a parameter
 characterizing the amplitude of 
 primordial non-Gaussianity, based on three-point correlations in the
 temperature and polarization anisotropies of the 
 cosmic microwave background. Using the example of the Planck mission
 our estimator is  faster by factors of 
 order $10^6$ than the estimator described by Babich     and    Zaldarriaga
 ~\citep{BZ04}, and yet provides essentially identical error bars. 

 The
 speed of our estimator allows us to study its statistical properties
 using Monte Carlo simulations. We have explored the effects of
 instrument noise (assuming homogeneous noise), 
 finite resolution, as well as sky cut. 
We conclude that our fast estimator is robust to these
 effects and extracts information optimally when compared to the
 Cramer Rao bound, in the limit of homogeneous noise.

We have uncovered a potential systematic effect that is important for
instruments measuring polarization with extremely high signal-to-noise on large scales.
The inevitable removal of contaminated portions of the sky causes any
estimator based on the pseudo-bispectrum to be contaminated by
mode-to-mode couplings at low $\ell$. We have demonstrated that by simply
excluding low $\ell$  polarization modes from the analysis removes
this systematic error with negligible information loss.

It has been shown that inhomogeneous noise causes cubic estimators based
on the psudo-bispectrum with a flat weighting to be significantly
suboptimal~\citep{nong_wmap}. A partial solution to this problem has
been found by \citet{creminelli_wmap1,creminelli_wmap2}, where a linear
piece has been added to the estimator in addition to the cubic piece. It
should be straightforward to apply their method to our estimator.

Finally, our reconstruction approach may be extended to find
fast estimators  for higher order statistics, for  example trispectrum
based estimators of $f_{NL}$~\citep{Kogo_Komatsu_06}. This is the subject of ongoing work.

\acknowledgments We acknowledge stimulating discussions with Michele
Liguori. EK acknowledges support from an Alfred P. Sloan Fellowship.
BDW acknowledges support from the Stephen Hawking Endowment
for Cosmological Research. Some of the  results in this
paper  have  been derived  using  the CMBFAST package by Uros
Seljak and Matias  Zaldarriaga \citep{5} and the  HEALPix package \citep{Healpix}. This  work  was  partially  supported by  the
National Center for Supercomputing Applications under TG-MCA04T015 and
by University of Illinois. We also utilized the Teragrid Cluster (www.teragrid.org)
at NCSA.  BDW and APSY's work is partially supported by  NSF AST O5-07676 and NASA JPL
  subcontract 1236748.

\bibliographystyle{apj}

\section{Derivation of Fast Cubic Estimator}
We  derive an  expectation  value  of the  cubic  statistics given  by
Eq. \ref{s_prim}
\begin{eqnarray}
\langle  \hat{S}_{prim}
 \rangle=\frac{1}{f_{sky}}\int  r^2dr \int  d^2\hat{n} \langle
B(\hat{n},r) B(\hat{n},r)A (\hat{n},r)
\rangle
\end{eqnarray} 
using the form of B and A as given by Eq.~\ref{B} and~\ref{A}
\begin{eqnarray}
\langle \hat{S}_{prim} \rangle =\frac{1}{f_{sky}}\sum_{i j k p q r} \sum_{\ell_1 \ell_2
\ell_3}     \sum_{m_1     m_2    m_3}(C^{-1})^{ip}_{\ell_1}(C^{-1})^{j
q}_{\ell_2}(C^{-1})^{k      r}_{\ell_3}      \langle     a^{i}_{\ell_1
m_1}a^{j}_{\ell_2       m_2}a^{k}_{\ell_3      m_3}
\rangle      \int
r^2dr\beta^p_{\ell_1}(r)\beta^{q}_{\ell_2}(r)\alpha^{r}_{\ell_3}(r) \nonumber 
\\
\int    d^2\hat{n}Y_{\ell_1    m_1}(\hat{n})Y_{\ell_2    m_2}(\hat{n})
Y_{\ell_3 m_3}(\hat{n})
\end{eqnarray}
which simplifies to
\begin{eqnarray}
\langle  \hat{S}_{prim}
  \rangle =\frac{1}{f_{sky}}\sum_{i  j  k  p q  r}\sum_{\ell_1
\ell_2              \ell_3}              I_{\ell_1              \ell_2
\ell_3}(C^{-1})^{ip}_{\ell_1}(C^{-1})^{j         q}_{\ell_2}(C^{-1})^{k
r}_{\ell_3}                                                        \int
r^2dr\beta^p_{\ell_1}(r)\beta^{q}_{\ell_2}(r)\alpha^{r}_{\ell_3}(r)
B^{i j k}_{\ell_1 \ell_2 \ell_3 }
\end{eqnarray}
where
\begin{eqnarray} 
I_{\ell_1 \ell_2  \ell_3}=\sqrt{\frac{(2\ell_1 +1)(2\ell_2 +1)(2\ell_3
+1)}{4\pi}}\left(\begin{array}{ccc} \ell_1 & \ell_2 & \ell_3\\ 0 & 0 &
0\end{array}\right),
\end{eqnarray}and
\begin{eqnarray}
B^{i  j   k}_{\ell_1  \ell_2   \ell_3  }=\sum_{m_1  m_2   m_3}  \left(
\begin{array}{ccc}   \ell_1  &  \ell_2   &  \ell_3\\   m_1  &   m_2  &
m_3\end{array}\right)B^{i   j  k}_{\ell_1   \ell_2   \ell_3  m_1   m_2
m_3}\end{eqnarray} is the angular  bispectrum, and $ B^{i j k}_{\ell_1
\ell_2  \ell_3 m_1  m_2  m_3}=\langle a^{i}_{\ell_1  m_1}a^{j}_{\ell_2
m_2}a^{k}_{\ell_3  m_3} 
\rangle $is  the CMB  bispectrum and  can be
averaged   as   above  due   to   isotropy.    In  deriving   $\langle
\hat{S}_{prim}
\rangle $ we have also used:
\begin{eqnarray}
\int    d^2\hat{n}Y_{\ell_1    m_1}(\hat{n})Y_{\ell_2    m_2}(\hat{n})
Y_{\ell_3     m_3}(\hat{n})     =I_{\ell_1    \ell_2     \ell_3}\left(
\begin{array}{ccc}   \ell_1  &  \ell_2   &  \ell_3\\   m_1  &   m_2  &
m_3\end{array}\right)\end{eqnarray} The  theoretical primordial angular
bispectrum can be written as :
\begin{equation}
B^{i j k}_{\ell_1 \ell_2 \ell_3 }=f_{NL} B^{p q r,prim}_{\ell_1 \ell_2
\ell_3 }
\end{equation}
where
\begin{eqnarray}
B^{p q r,prim}_{\ell_1 \ell_2  \ell_3 }=2 I_{\ell_1 \ell_2 \ell_3}\int
r^2dr      \lbrack     \beta^p_{\ell_1}(r)\beta^{q}_{\ell_2}(r)\alpha^
{r}_{\ell_3}(r)                                                       +
\beta^{r}_{\ell_3}(r)\beta^{p}_{\ell_1}(r)\alpha^{q}_{\ell_2}(r)      +
\beta^{q}_{\ell_2}(r)\beta^{r}_{\ell_3}(r)\alpha^{p}_{\ell_1}(r)
\rbrack  \end{eqnarray}  Using  the  above  form  of  the  theoretical
bispectrum, $\langle \hat{S}_{prim} \rangle$ further simplifies to
\begin{eqnarray}
\langle \hat{S}_{prim}
 \rangle=\frac{1}{f_{sky}}\sum_{i j k p q r}\sum_{\ell_1 \ell_2
\ell_3}    \!    &\frac{1}{6}&\!B^{p    q    r,prim}_{\ell_1    \ell_2
\ell_3}(r)(C^{-1})^{ip}_{\ell_1}(C^{-1})^{j   q}_{\ell_2}  (C^{-1})^{k
r}_{\ell_3} B^{i j k}_{\ell_1 \ell_2 \ell_3 }.
\label{eq:est2}
\end{eqnarray}
Now since  $\sum_{\ell_1 \ell_2 \ell_3}  = 6\sum_{\ell_1 \le
\ell_2  \le  \ell_3} \frac{1}{\Delta_{\ell_1 \ell_2 \ell_3}}$, where $\Delta_{\ell_1  \ell_2
\ell_3}$, which  is 1  when $\ell_1 \neq  \ell_2 \neq \ell_3$,  6 when
$\ell_1 = \ell_2 = \ell_3$, and 2 otherwise.
\begin{eqnarray}
\langle \hat{S}_{prim}
 \rangle = \frac{1}{f_{sky}}\sum_{i j k p q r}\sum_{\ell_1
 \le   \ell_2  \le   \ell_3}  \frac{1}{\Delta_{\ell_1 \ell_2 \ell_3}}f_{NL}B^{p   q   r,prim}_{\ell_1  \ell_2
 \ell_3}(C^{-1})^{ip}_{\ell_1}(C^{-1})^{j    q}_{\ell_2}   (C^{-1})^{k
 r}_{\ell_3} B^{i j k,prim}_{\ell_1 \ell_2 \ell_3 }
\end{eqnarray}

\section{Proof of Covariance matrix}
In  this  appendix  we  prove  the  equivalence   between  the  optimal
estimator given by \citep{BZ04} and our fast estimator 
Eqn. (\ref{eq:est2}) 
\begin{eqnarray}
 \sum_{ijkpqr}\sum_{\ell_1 \ell_2 \ell_3} \frac{1}{6} B^{p  q r,prim}_{\ell_1
\ell_2 \ell_3}(C^{-1})^{ip}_{\ell_1}(C^{-1})^{j q}_{\ell_2}(C^{-1})^{k
r}_{\ell_3}    B^{i     j    k,prim}_{\ell_1    \ell_2     \ell_3    }
=\sum_{\alpha \beta}\sum_{\ell_1 \ell_2 \ell_3}
B^{\alpha,prim}_{\ell_1\ell_2\ell_3}({\bf Cov}^{-1})_{\alpha,\beta}
B^{\beta,prim}_{\ell_1\ell_2\ell_3},
\end{eqnarray}
This is analogous  to the  temperature-only case \citep{KSW05}.
On the left  hand side $ijk$  and $pqr$
run     over      all      the     8      possible   ordered combinations
$\{TTT,TTE,TET,ETT,TEE,ETE,EET,EEE\}$, while for the estimator
on  the  right  hand  side  $\alpha$  and $\beta$  run  only over the four
unordered combinations $\{TTT,TTE,TEE,EEE\}$. 

We  prove the equivalence between
the optimal estimator and the  fast estimator for only one combination
of $\ell_1,$ $\ell_2,$ and $\ell_3$, as  the proof is same for all the
combinations.The covariance matrix {\bf Cov} is obtained in terms of $C_\ell^{TT}$,
$C_\ell^{EE}$,  and $C_\ell^{TE}$  (as  in equation  7  in Babich  and
Zaldarriaga (2004)) by applying Wick's theorem,
\begin{eqnarray}
\mathbf{Cov}_{ijk,pqr}=    C^{ir}    C^{jq}    C^{kp}+C^{iq}    C^{jr}
C^{kp}+C^{ir}  C^{jp}  C^{kq}+C^{ip}   C^{jr}  C^{kq}+  C^{iq}  C^{jp}
C^{kr}+C^{ip} C^{jq} C^{kr}.
\end{eqnarray}
The covariance matrix above has the
form $  {\bf Cov}_{\alpha\,\beta},$ where $\alpha$,  $\beta$, run over
$\{TTT,TTE,TEE,EEE\}$, which after simplification gives:
\begin{footnotesize}
\begin{eqnarray*}\label{CovarianceProof}
{\bf Cov}^{-1}_{\alpha\,\beta} =\frac{1}{6}\left(
\begin{array}{llll}
 {\hat  C^{TT}}  {\hat  C^{TT}}{\hat  C^{TT}}&\!  3{\hat  C^{TT}}{\hat
C^{TT}}\hat  C^{TE} &\!  3 \hat  C^{TT}{\hat C^{TE}}  {\hat C^{TE}}&\!
{\hat  C^{TE}}  {\hat  C^{TE}}{\hat  C^{TE}}\\  3{\hat  C^{TT}}  {\hat
C^{TT}}{\hat    C^{TE}}    &\!     3{\hat   C^{EE}}    {\hat
C^{TT}}{\hat         C^{TT}}+6{\hat        C^{TE}}{\hat
C^{TE}}{\hat  C^{TT}}  &\!   3  {\hat  C^{TE}}{\hat  C^{TE}}{\hat
C^{TE}}+6\hat  C^{TT}\hat C^{EE}\hat C^{TE}  &\! 3  {\hat C^{TE}}{\hat
C^{TE}}\hat  C^{EE} \\  3{\hat C^{TT}}{\hat  C^{TE}}{\hat  C^{TE}} &\!
3{\hat    C^{TE}}{\hat   C^{TE}}{\hat    C^{TE}}+6{\hat   C^{TT}}{\hat
C^{EE}}{\hat C^{TE}} &\!   6{\hat C^{EE}} {\hat C^{TE}}{\hat C^{TE}}+3
{\hat  C^{TT}}{\hat  C^{EE}}   {\hat  C^{EE}}&\!  3{\hat  C^{TE}}{\hat
C^{EE}}  {\hat C^{EE}}\\ {\hat  C^{TE}} {\hat  C^{TE}}{\hat C^{TE}}&\!
3{\hat  C^{TE}}{\hat  C^{TE}}{\hat  C^{EE}}  &\!  3{\hat  C^{TE}}{\hat
C^{EE}} {\hat C^{EE}}&\!  {\hat C^{EE}}{\hat C^{EE}}{\hat C^{EE}}
\end{array}
\right),
\end{eqnarray*} 
\end{footnotesize}
where $\hat C^{XY} \equiv (C^{-1})^{XY}$  are the $XY$ elements of the
matrix                    $\left(                   \begin{array}{ccc}
C^{TT}&C^{TE}\\C^{TE}&C^{EE}\\\end{array} \right)^{-1}\,  $. The above
matrix  is nothing  but  $\frac{1}{6} (C^{-1})^{ip}(C^{-1})^{j q}(C^{-1})^{k  r}$,
after the additional permutations with fixed $\alpha$ and $\beta$
are summed up. For example, for $\alpha=TTE$ the $ijk$ index runs over
the set
$\{TTE,TET,ETT\}$. This completes the proof.

\end{document}